\documentclass[showpacs,twocolumn,amssymb,prl,aps]{revtex4}
\usepackage{graphicx,epsfig}
\usepackage{color}

\begin{document}
\title{Paranematic-to-nematic ordering of a binary mixture of rod-like liquid crystals confined in cylindrical nanochannels}
\author{Sylwia~Ca{\l}us$^{1}$ }
\author{Beata~Jab{\l}o{\' n}ska$^{2}$}
\author{Mark~Busch$^{3}$}
\author{Daniel~Rau$^4$}
\author{Patrick~Huber$^{3,4}$}
\email[E-mail: ]{patrick.huber@tuhh.de}
\author{Andriy~V.~Kityk$^{1}$ }
\email[E-mail: ]{andriy.kityk@univie.ac.at}

\affiliation{$^1$Faculty of Electrical Engineering, Czestochowa University of Technology, 42-200 Czestochowa, Poland\\
$^2$Faculty of Environmental Engineering and Biotechnology, Czestochowa University of Technology, 42-200 Czestochowa, Poland\\
$^3$ Materials Physics and Technology, Hamburg University of Technology (TUHH), D-21073 Hamburg-Harburg, Germany\\
$^4$ FR 7.2 Experimental Physics, Saarland University, D-66123 Saarbr\"ucken, Germany}

\date{\today}

\begin{abstract}
We explore the optical birefringence of the nematic binary mixtures 6CB$_{1-x}$7CB$_x$ ($0~\le~x~\le~1$) imbibed into parallel-aligned nanochannels of mesoporous alumina and silica membranes for channel radii of $3.4~\le~R~\le 21.0$ nm. The results are compared with the bulk behavior and analyzed with a Landau-de-Gennes model. Depending on the channel radius the nematic ordering in the cylindrical nanochannels evolves either discontinuously (subcritical regime, nematic ordering field $\sigma<1/2$) or continuously (overcritical regime, $\sigma>1/2$), but in both cases  with a characteristic paranematic precursor behavior. The strength of the ordering field, imposed by the channel walls, and the magnitude of quenched disorder varies linearly with the mole fraction $x$ and scales inversely proportional with $R$ for channel radii larger than 4 nm. The critical pore radius, $R_c$,  separating a continuous from a discontinuous paranematic-to-nematic evolution, varies linearly with $x$ and differs negligibly between the silica and alumina membranes. We find no hints of preferred adsorption of one species at the channels walls. By contrast, a linear variation of the nematic-to-paranematic transition point $T_{\rm PN}$ and of the nematic ordering field $\sigma$ vs. $x$ suggest that the binary mixtures of cyanobiphenyls 6CB and 7CB keep their homogeneous bulk stoichiometry also in nanoconfinement, at least for channel diameters larger than $\sim$7~nm.
\end{abstract}

\maketitle

\section{1. Introduction}
For decades binary mixtures of liquid crystals (LC) have been attracting not only a considerable fundamental interest, but have been employed in various functional applications. Particularly, low molecular weight nematics, used in display technology, are routinely mixed to reduce the melting temperature and thus to increase the operating range of LC displays down to substantially lower temperatures \cite{Yeh}. In the absence of chemical reactions the bulk properties of many nematic binary mixtures, especially those which are composed of similar LC components, are represented as a weighted sum of the individual properties of the pure constituents. Hence their physical characteristics, such as dielectric constants and anisotropy, elastic constants, birefringence and viscosity appear to be well predictable and/or tunable. This allows for an optimisation of basic parameters of LC displays \cite{Kim,Gauza}.

\begin{figure}[tbp]
\epsfig{file=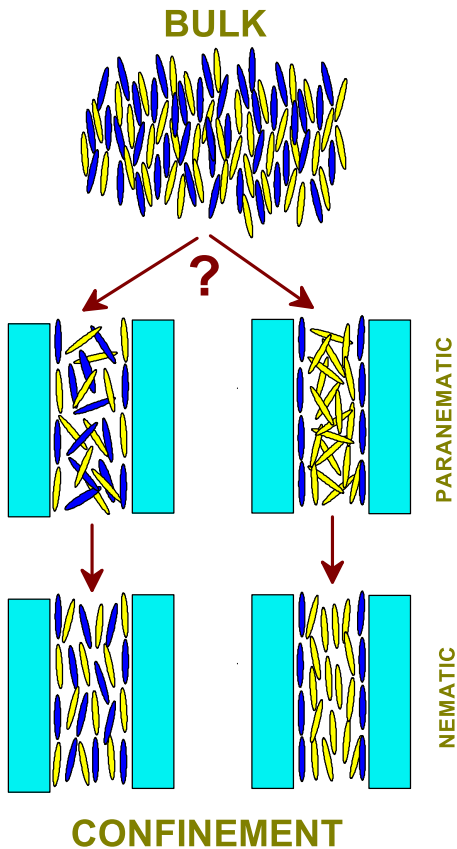, angle=0,
width=0.7\columnwidth}\caption{\footnotesize (color online). Possible arrangements of a binary nematic mixture in cylindrical nanochannels. The two species of nematogen molecules are marked in blue and yellow. In both cases the nematic ordering is enforced by the cylindrical geometrical constraint (the so-called nematic ordering field). However, the structural arrangement may be different depending on whether the stoichiometry remains homogeneous over the pore volume (left) or is subjected to preferred adsorption of one species at the channel walls (right).} \label{fig1}
\end{figure}

Recent trends in electronic and optoelectronic miniaturization shift many applications towards the nanoscale or rely on nanocomposits. Nematic binary mixtures embedded into nanoporous substrates or more generally spoken confined in nano-scale geometries represent therefore material systems which increasingly attract interest both with regard to fundamental scientific \cite{Steuer2004,Binder2008,Schoenhals2010,Mazza2010,MD,Jasiurkowska2012} but also applied aspects \cite{Grigoriadis2011}. The basic problems are not limited here just to specific properties of the guest materials or to a characteristic pore size of the nanoporous host substrates, which may vary between few and hundreds nanometers. They strongly depend on the guest-host interfacial interactions as well as the actual geometry of the pores and/or the pore network.  The interfacial interactions are transferred through subsequent neighboring molecular layers modifying the core region of a pore filling as well \cite{Kityk2008}. Due to their decaying character the properties of the guest material in particular and the nanocomposite as a whole are expected to deviate considerably from the bulk state.

In the case of one component LCs confinement effects have been experimentally explored either by embedding of LCs in porous substrates
\cite{Bellini,Iannac,Iannac1,Guegan,Sihai,Kralj,Kutnjak,Kutnjak1, Schoenhals2010, Grigoriadis2011, Kityk2014} or by dispersion of small particles in LC fluids \cite{Iannac2,Kutnjak2}. The resulting properties appear to be strongly dependent on the average pore size, types of the embedded LC, material of the porous matrix as well as eventual surface-treatment of the pore walls. For instance, substrates with sufficiently small pore sizes render the isotropic-to-nematic phase transition gradual.

Accordingly, even far above the nematic state there exists a weak residual nematic ordering, a so-called paranematic state, which is preferentially located in the interface region. This phenomenology is similar to the one encountered for a ferromagnetic system subjected to an external magnetic field, which replaces a discrete phase transition by a continuous evolution of the order parameter \cite{Iannac}. 

Referring to and building up on, the pioneering theoretical work of Sheng, Poniewierski, Sluckin \cite{Sheng1976, Poniewierski1987} as well as  experimental work by Yokoyama\cite{Yokoyama1988} on LCs in semi-infinite, planar confinement, Kutnjak, Kralj, Lahajnar, and Zumer developed a phenomenological approach \cite{Kutnjak,Kutnjak1} (hereafter denoted as KKLZ model) for the isotropic-to-nematic transition in pore geometries by introducing a so-called \emph{nematic ordering field} $\sigma$ into the Landau-de Gennes free energy, which couples bilinearly with the nematic order parameter. According to their concept $\sigma$  is proportional to the anchoring strength $W$ and inversely proportional to the pore radius $R$. In other words, it is defined by the physico-chemical interactions at the interface and geometrical constraints.

Experimentally, paranematic behavior in nano channels has been directly demonstrated in NMR \cite{Kralj} and high resolution optical birefringence \cite{Kityk2008,Kityk1} studies on rod-like cyanobiphenyl nematogens. A careful analysis of the measured excess birefringence on nematic 5CB and 6CB embedded into parallel aligned cylindrical nanochannels of porous silicon and silica membranes \cite{Calus} has indeed proved the $R^{-1}$-dependence of the nematic ordering field, $\sigma$, at least for pore radii larger than 4.7 nm.

The objective of this work was to study the molecular ordering in the nematic mixtures composed of the nematogen LCs, 6CB (4-cyano-4'-hexylbiphenyl) and 7CB (4-Cyano-4'-heptylbiphenyl), confined into radius-controlled arrays of parallel-aligned nanochannels in porous silica ($p$SiO$_2$, 3.4$\le R \le$ 6.4 nm ) and porous alumina ($p$Al$_2$O$_3$, 10$\le R \le$ 21 nm ) membranes. The nematic ordering is explored by high resolution optical birefringence and the experimental results are compared with the bulk behavior and analyzed within the KKLZ approach.

The choice of these low-molecular LC compounds is motivated by the two following reasons:

 (i) Both pure cyanobiphenyls are characterized by identical phase diagrams. During cooling, before solidification, they exhibit only a single first order phase transition from the isotropic (I) to the nematic (N) phase at $T_{IN}\approx$ 302.4 K for 6CB \cite{Janik} and $T_{IN}\approx$315.5 K (7CB) \cite{Dn};

 (ii) For relatively short molecules, as 6CB or 7CB, the strength of the first-order character of the isotropic-to-nematic transition is not expected to be affected by the confinement as has been ascertained in recent theoretical studies \cite{Kralj10,Kralj11}. This substantially simplifies the theoretical analysis based on the KKLZ approach. We assume that this statement remains valid not only for the pure compounds, 6CB and/or 7CB, but also for their mixtures.

Whereas the main goals of this study are an experimental exploration and a fundamental understanding of the phase transition behavior of the spatially confined, mixed system, an additional question concerns a possible demixing in the nanochannels. Depending on the mutual interactions of the LC components and the LC component/wall interactions inhomogeneous concentrations within the pores may result, similarly as it has been theoretically predicted \cite{Gelb1999,Woywod2003, Woywod2005} and experimentally found for simple liquids \cite{Rother2004, Schemmel2005,Lefort2011}. Two possible scenarios are drawn in Fig.~1. Here the rod-like molecules of different types are marked by blue and yellow colors, respectively. In both cases the nematic ordering along the long channel axis is enforced by the cylindrical geometry. However, the molecular ordering may depend on whether the local stoichiometry remains constant over the channel volume (and corresponds to that of the as prepared bulk state), see left panel in Fig.~1, or is affected by a separation of the components caused by a preferred adsorption of one constituent at the channel walls, as sketched in the right panel of Fig.~1. Therefore, a phase separation in the channels would significantly hamper the phenomenological description. Fortunately, as we demonstrate below, we found no hints for such a phase separation.

\section{2. Experimental}

Nematogen LCs 6CB and 7CB have been purchased from Merck.  Binary mixtures, 6CB$_{1-x}$7CB$_x$, where $x$ corresponds to the mole fraction of the 7CB component, were prepared at room temperature by weighting the pure components in the selected proportions ($(1-x)/x=$ 1.0/0, 0.839/0.161, 0.676/0.324, 0.526/0.474, 0.328/0.672, 0.198/0.802, 0/1.0).  The components were first mixed, then heated up slightly above the clearing point and annealed for a period of several days to achieve homogeneous mixtures.

The porous silica membranes, $p$SiO$_2$, have been prepared in two consequent steps. In the first step, the porous silicon ($p$Si) membranes have been obtained by electrochemical anodic etching of highly $p$-doped,  $\langle$100$\rangle$ oriented silicon wafers. In the second step, the free standing $p$Si membranes have been subjected to further thermal oxidation for 12 h at $T$=800 $^o$C under standard atmosphere. By applying different etching conditions we have obtained  $p$SiO$_2$ membranes with average pore radii 3.4$\pm$0.3 nm (porosity $P$=13\%, thickness $h=$110 $\mu$m), 4.5$\pm$0.4 nm ($P$=30\%, $h=$170 $\mu$m) and  6.4$\pm$0.5 nm ($P$=45\%, $h=$290 $\mu$m) as verified by recording of volumetric N$_2$-sorption isotherms at $T$=77 K.

The porous alumina membranes, ($p$Al$_2$O$_3$), were purchased from Smart Membranes GmbH (Halle, Germany). Regardless of the pore size specified by the company, the pore diameter has been verified  by volumetric N$_2$-sorption isotherms at $T$ = 77~K yielding the following average channel radii: $R = 21.0 \pm$2.0 nm ($P$=24\%, $h=$100~ $\mu$m), 15.5$\pm$1.5 nm ($P=$17\%, $h=$100 $\mu$m) and 10.0$\pm$0.7 nm ($P=$16\%, $h=$90 $\mu$m).

\begin{figure}[tbp]
\begin{center}
  \epsfig{file=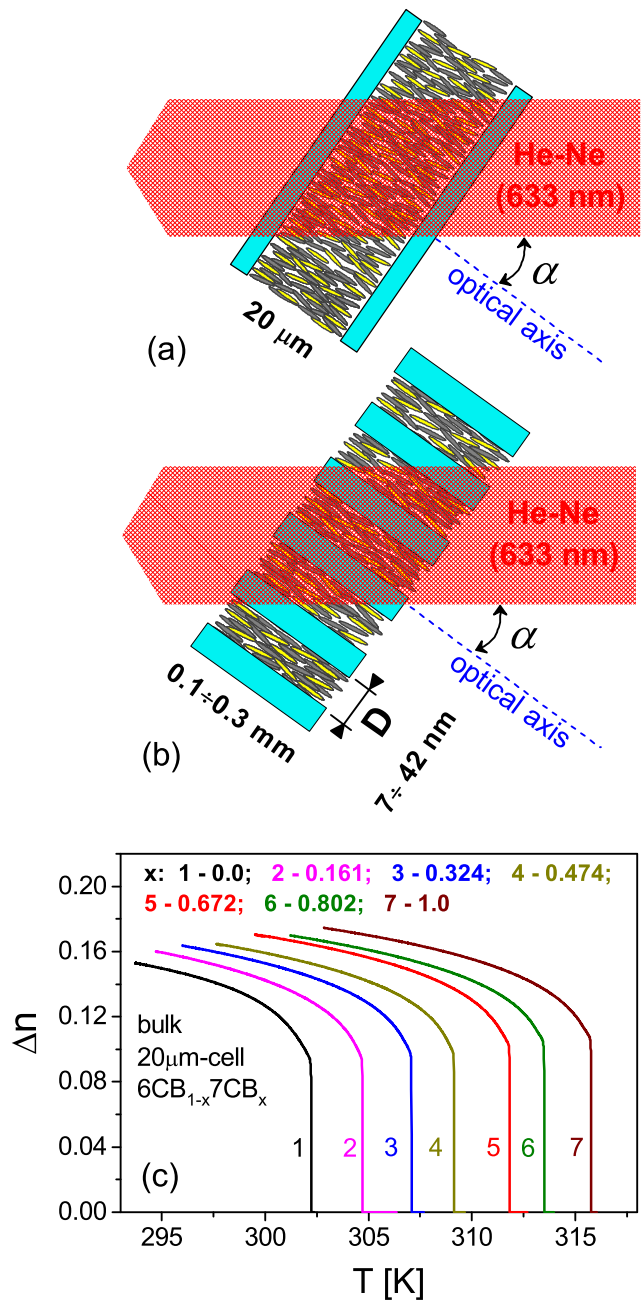, angle=0, width=0.99\columnwidth}
  \caption{\footnotesize (color online). Schematic sketch of homeotropic alignment of rod-like molecules in a cell of parallel glass plates (a) and parallel axial molecular alignment in cylindrical nanochannels (b). In both cases the refractive index anisotropy is characterized by a single optical axis being perpendicular to the sample surface. In order to measure the birefringence the samples were turned out with respect to the incident He-Ne laser beam ($\lambda=$ 633 nm) by an angle, $\alpha \sim$ 40-45 deg as indicated in the figure. In panel (c) the bulk optical birefringence of  the nematic LCs, 6CB ($x=$0) and 7CB ($x=$1) as well as a series of their binary mixtures 6CB$_{1-x}$7CB$_x$ ($x=$ 0.161, 0.324, 0.474, 0.672, 0.802) is shown as a function of temperature, as measured during heating in the parallel glass 20 $\mu$m-cell.} \label{fig2}
\end{center}
\end{figure}

First, the mesoporous membranes were cleaned in boiling acetone and chloromethane ($\sim$1 hour in each solvent) several times and then dried at about 100 $^o$C for a few hours. Then they were completely filled by capillary action (spontaneous imbibition) of the nematic mixture with a certain mole fraction $x$ in the isotropic phase  \cite{Gruener2010} at about 325 K. To avoid uncertainties related to uncontrolled deviations of pore diameters and/or porosity in the membranes of the same type the optical polarimetric measurements have been performed using the same set of nanoporous membranes with pore radii $R$ as specified above, refilling them for subsequent measurements by nematic mixtures of a different mole fraction, $x$.  The membranes were carefully cleaned from the pore filling remaining from a previous measurement by the procedure outlined above, i.e. by immersing them into boiling acetone for about 1 hour and repeating this 4-5 times with a fresh acetone solvent. As we carefully checked, such a procedure provides an overall behavior of the optical birefringence which is fully reproducible, independently on the filling history of the samples, that is on the sequence of nematic binary mixtures investigated.

For the bulk measurements the sample cells, made of silanized parallel glass plates ($d$=20 $\mu$m), have been filled by the LCs providing an homeotropic alignment, as it is sketched in Fig.~2a. The untreated surface of silica  pores, on the other hand, enforce planar anchoring without a preferred lateral direction within the plane \cite{Drevensek2003, zanoni1}, whereas the elongated geometry of cylindrical pores yields, additionally, a preferred orientation of molecules along the pores axis resulting in a so-called parallel axial alignment, see Fig.~2b. The parallel axial alignment is also typical of the alumina pores as it follows from the measurements provided below. This is not too surprising, if one considers that the alumina surface, similar to the silica one, is covered by OH-groups, which in fact enforce a planar type of anchoring.

The molecular alignment both in the bulk cell and in the nanoporous membranes results in a refractive index anisotropy which is characterized by a single optical axis, which is perpendicular to the sample surface, as it is shown in Fig.~2a and 2b. A change in the molecular alignment results in a change of the optical birefringence. This change was measured by a high-resolution polarimetry setup, which employs a photoelastic modulator and a dual lock-in detection system, as described in Ref.\cite{Kityk2008}. It provides an accuracy of the optical retardation measurements better than 5$\cdot$10$^{-3}$ deg. Note that the sample was tilted out with respect to the optical axis by an angle, $\alpha \sim$ 40-45 deg - see Fig.~\ref{fig2}.

\begin{figure*}[tbp]
\epsfig{file=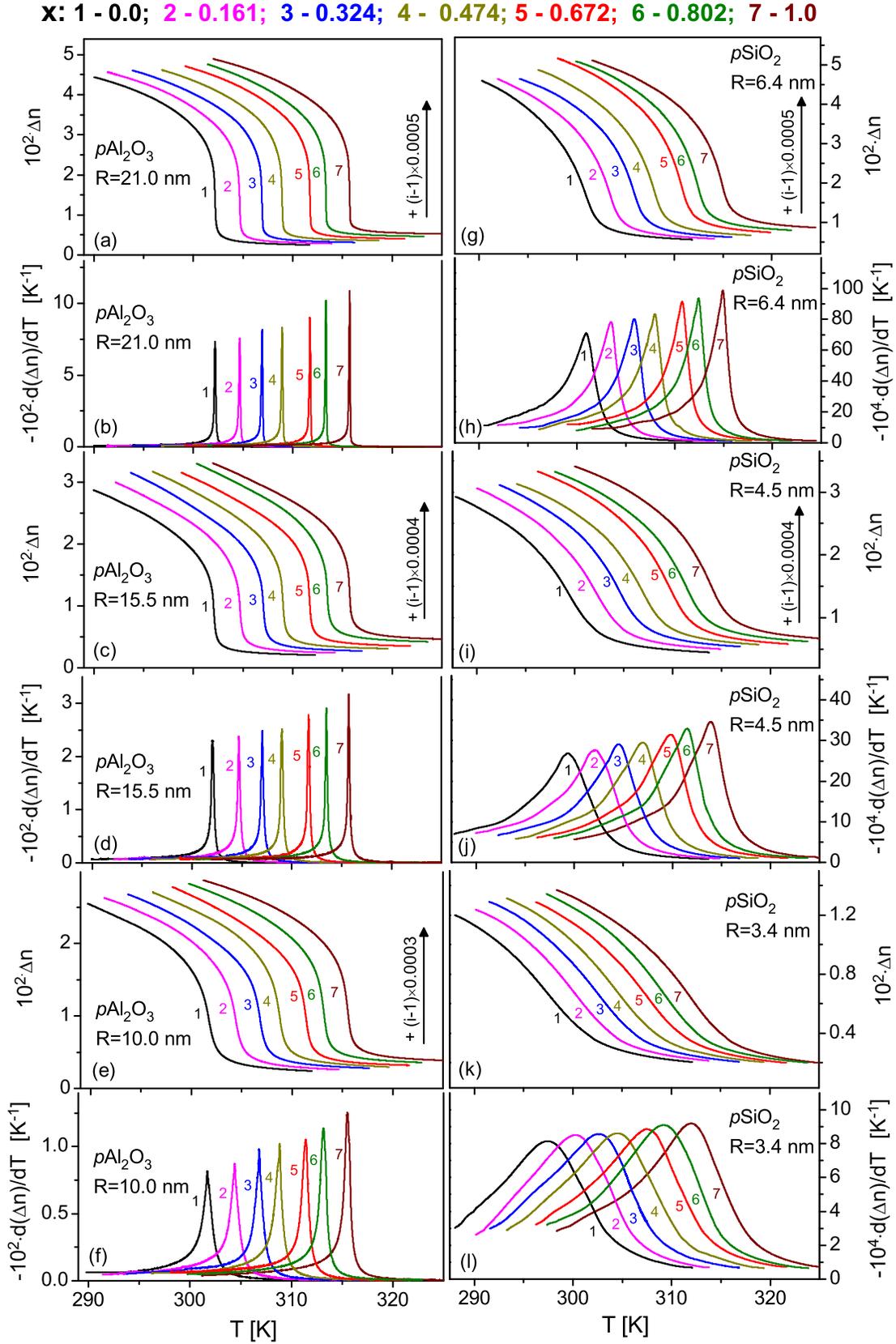, angle=0,width=1.7\columnwidth}
 \caption{\footnotesize (color online). Measured optical birefringence $\Delta n$ vs temperature $T$ for the pure nematic LCs, 6CB and 7CB  and for a series of their binary mixtures 6CB$_{1-x}$7CB$_x$ ($x=$ 0.161, 0.324, 0.474, 0.672, 0.802) embedded into parallel-aligned nanochannels of porous alumina membranes $p$Al$_2$O$_3$: $R=21.0$ nm [panel (a)], $R=15.5$ nm [panel (c)], $R=10.0$ nm [panel (e)] and porous silica membranes $p$SiO$_2$: $R=6.4$ nm [panel (g)], $R=4.5$ nm [panel (i)], $R=3.4$ nm [panel (k)]. For clarity the curves in panels (a),(c),(e),(g) and (i) are incrementally shifted upwards, see the arrows with labeled increment values, where $i$ refers to the number of the shifted curve. In the panels (b),(d),(f),(h),(j) and (l) the temperature derivatives, $-d(\Delta n)/dT$, vs $T$, extracted from the data presented in panels (a),(c),(e),(g),(i) and (k) (see labels) are depicted.}
 \label{fig3}
\end{figure*}

\section{3. Results and discussion}

In the bulk nematic state the optical birefringence changes linearly with the orientational order parameter $Q = \frac{1}{2}\langle 3\cos^2\theta-1\rangle$, where $\theta$ is the angle between the long axis of a single molecule and a direction of preferred orientation of that axis, the director \cite{Kumar}. The brackets denote an averaging over all molecules under consideration.

\begin{figure}[tbp]
\begin{center}
\epsfig{file=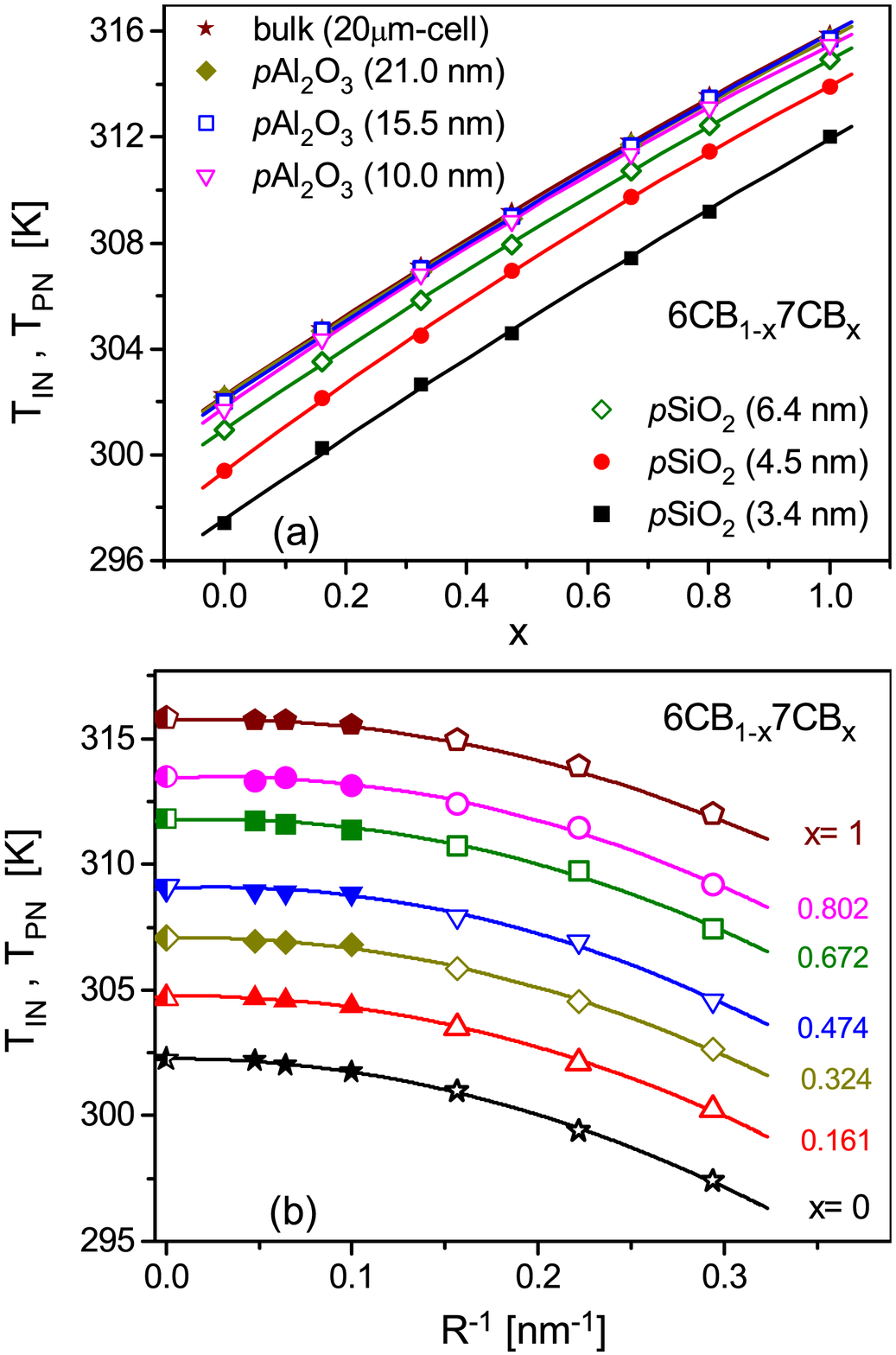, angle=0, width=0.99\columnwidth}
\caption{\footnotesize (color online). Panel (a): Temperatures of the nematic-to-isotropic bulk transition, $T_{\rm IN}$ and nematic-to-paranematic transition, $T_{\rm PN}$ vs mole fraction $x$ of the binary mixture  6CB$_{1-x}$7CB$_x$. The membrane types and the corresponding mean channel radii $R$ are listed as labels. Panel (b): $T_{\rm PN}$ vs the inverse channel radius, $R^{-1}>0$, for LCs 6CB ($x=$0.0) and 7CB ($x=$1.0) and a series of their binary nematic mixtures 6CB$_{1-x}$7CB$_x$, see labeled $x$-value. Semiopen symbols, located at $R^{-1}=$0, refer to the bulk transition temperature, $T_{\rm IN}$. Solid and open symbols correspond to alumina and silica membranes, respectively.  $T_{\rm IN}(x)$ is determined from the jump position in the $\Delta n(T)$-dependences, see Fig.~2(c). The values of $T_{\rm PN}(x,R)$ are determined as the position of the derivative peaks shown in Fig.~3 [Panels: (b),(d),(f),(h),(j) and (l)].} \label{fig4}
\end{center}
\end{figure}

In Fig. \ref{fig2}(c) the temperature dependences of the bulk optical birefringence of the pure nematic LCs, 6CB ($x=$0) and 7CB ($x=$1), as well as a series of their binary mixtures 6CB$_{1-x}$7CB$_x$ ($x=$ 0.161, 0.324, 0.474, 0.672, 0.802) measured during heating in the parallel glass 20 $\mu$m-cell are depicted. As usual, the N-I transition is characterized by a jump-like vanishing of the optical birefringence in accordance with its first order character. Such behavior is observed both for pure 6CB and pure 7CB as well as for their mixtures. We explicitly emphasize that there is no indication of a smearing of the phase transition in the binary mixtures, which documents their homogeneity. During cooling the measured birefringence exhibits practically an identical behavior with a small temperature hysteresis (less than 0.1 K) of the I-N transition. Therefore, we avoid a presentation of cooling scans in the following. For pure 6CB and 7CB the observed phase transition temperatures, $T_{\rm IN}$, are 302.25 K and 315.7 K, respectively, i.e. they agree within the experimental error with the values reported in the literature \cite{Janik,Dn}. While the mole fraction number $x$ varies from 0.0 to 1.0, the phase transition point $T_{\rm IN}$ in the binary nematic mixtures, 6CB$_{1-x}$7CB$_x$,  shifts almost linearly between these temperatures, see Fig.~\ref{fig4}(a).

The orientational order inside the channels results in an excess birefringence, $\Delta n^+ \propto Q$, which scales with the porosity of the membrane and appears on the background of the geometric birefringence, $\Delta n_g$, being a characteristic feature of membranes with parallel nanochannels \cite{Kityk2}. Accordingly, a simple linear relation between optical birefringence and $Q$ is conserved for the LC-based nanocomposite, provided the  geometrical birefringence is subtracted. 

For the entirely filled membranes the geometric birefringence is relatively small, because of the similar refractive indices of the host matrix and guest LCs or their binary mixtures. Nevertheless, it exhibits a weak temperature behavior, referred hereafter as "thermooptic" changes. They originate in the varying refractive indices of the host and guest materials as well as changes of the porosity caused by the thermal expansion/contraction of the porous matrix. It can be appropriately considered by a linear dependence in the fitting procedure \cite{Calus}.

In Fig.~3 we show the optical birefringence $\Delta n$ vs. temperature $T$ for the nematic LCs, 6CB and 7CB, and a series of their binary mixtures 6CB$_{1-x}$7CB$_x$ ($x=$ 0.161, 0.324, 0.474, 0.672, 0.802) embedded into parallel-aligned nanochannels of porous alumina membranes $p$Al$_2$O$_3$: $R=21.0$ nm [panel (a)], $R=15.5$ nm [panel (c)], $R=10.0$ nm [panel (e)] and porous silica membranes $p$SiO$_2$: $R=6.4$ nm [panel (g)], $R=4.5$ nm [panel (i)], $R=3.4$ nm [panel (k)] during heating. Whereas for large pore radii, e.g. $R = $15.5 nm or 21.0 nm, the temperature dependences of $\Delta n$ are reminiscent of the bulk behavior, i.e. they still exhibit jump-like changes of the optical birefringence in the transition region, the transformation becomes more and more gradual with decreasing pore radii ($R \le$ 10 nm). According to the KKLZ model, the continuous and discontinuous behavior of $\Delta n(T)$ found for the small and large channels ($R \ge 15.5$ nm), resp., can be described by a sufficiently large or small nematic ordering field, $\sigma>\sigma_c=0.5$ and $\sigma<\sigma_c=0.5$ .

As outlined below a more detailed characterization of the observed behavior can be achieved by an appropriate fitting procedure. It allows one to distinguish between the subcritical ($\sigma<0.5$) and overcritical ($\sigma>0.5$) behavior. Independently of the criticality of the regime, there is a specific precursor behavior, typical of a residual nematic ordering, i.e. a paranematic state, observable for all confined systems. Hence, the birefringence evolution reflects a paranematic-to-nematic transition with characteristic temperatures marked as $T_{\rm PN}$. During heating the full (nematic order) below $T_{\rm PN}$ evolves into a partial collective orientational ordering (paranematic state) above $T_{\rm PN}$.

It is interesting to note, that the symmetry of both states are identical. Thus the P-N transformation does not represent a phase transition in a classical sense. Following the definition given in Ref.\cite{Huber2} $T_{\rm PN}$ can be defined as the temperature of fastest variations in $Q(T)$, that is the temperature corresponding to the maximum of the temperature derivative, $-d(\Delta n)/dT$. In panels (b),(d),(f),(h),(j) and (l) of Fig.~3 the temperature derivatives, $-d(\Delta n)/dT$ vs. $T$ are depicted, as calculated based on the data presented in panels (a),(c),(e),(g),(i) and (k) of Fig.~3. For a fixed mole fraction $x$ sharp derivative peaks observed at large pore radii considerably broaden with decreasing radii, e.g. compare Fig.~3(b) and Fig.~3(l). Evidently, such a peak evolution can be interpreted as the result of an increasing nematic ordering field, $\sigma$. Moreover, the peak height at a fixed $R$ rises with $x$, which qualitatively indicates a tendency for a decrease of the nematic ordering field and/or strengthening of the first order character of the paranematic-to-nematic transition with $x$. A more quantitative characterization in this respect will be given by a detailed fitting at a later stage of the manuscript.

\begin{figure*}[tbp]
\begin{center}
\epsfig{file=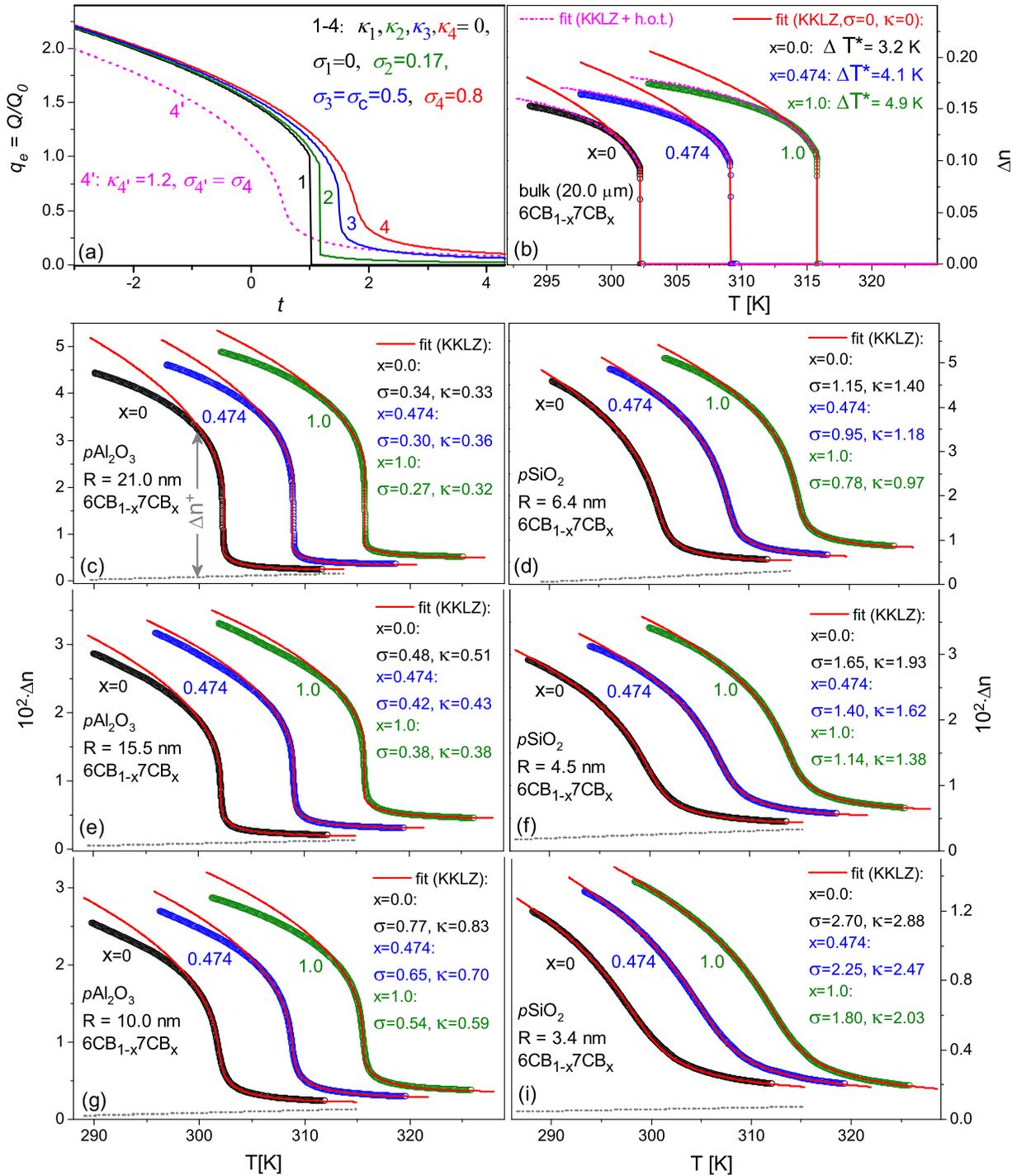, angle=0, width=1.9\columnwidth}
\caption{\footnotesize (color online). Analysis of the temperature behavior of the optical birefringence, $\Delta n(T)$, in the vicinity of the nematic-to-isotropic (paranematic) transition. Panel (a): Temperature behavior of the scaled equilibrium order parameter $q_e(t)$ for the bulk state (curve 1: $\sigma=0,\kappa=0$) and several confined geometries corresponding to subcritical (curve 2: $\sigma<\sigma_c,\kappa=0$), critical (curve 3: $\sigma=\sigma_c,\kappa=0$) and overcritical (curve 4: $\sigma>\sigma_c,\kappa=0$; curve 4$'$: $\sigma>\sigma_c,\kappa\ne0$) regimes ($\sigma_c=0.5$) as predicted within the KKLZ-approach. Panels (b)-(i): The optical birefringence of LCs 6CB ($x=$0), 7CB ($x=1.0$) and their binary mixtures 6CB$_{1-x}$7CB$_x$ ($x=$0.474) vs temperature $T$ in the bulk state [(b)] and imbibed into parallel-aligned nanochannels of porous alumina [(c),(e),(g)]  and silica [(d),(f),(i)] membranes. The channel radii, $R$, are specified by labels. The open symbols of black, blue or green colors (see labels) correspond to the measured birefringence as depicted in Fig.~3.  The solid lines (red online color) represent the best fits obtained within the KKLZ-model based on a free energy expansion (1). The base line corresponds to the thermooptic contribution (dash-dot line, gray online color). It is obtained within the same fitting procedure and is presented as an example for 6CB only. The extracted fit values, $\Delta T^*,\sigma$ and $\kappa$ are labeled in accordance with the corresponding curves. The full set of extracted fitting parameters (for all mole fractions $x$ studied in the present work) is presented in Table~1. The dash-dot lines (magenta color online) in Panel (b) are the calculated bulk birefringence using the free energy expansion completed by high order terms (up to 14th order).} \label{fig5}
\end{center}
\end{figure*}

The temperature $T_{\rm PN}$, determined from the derivative peak position, vs the mole fraction $x$ is shown in Fig.~4(a) for fixed pore radii $R$. In all cases $T_{\rm PN}(x)$ varies linearly between the paranematic-to-nematic transition points $T_{\rm PN}(R,x=0.0)$ and $T_{\rm PN}(R,x=1.0)$ of the pure LCs. For large pore radii, particularly $R=$ 21.0 nm or 15.5 nm, the $T_{\rm PN}(x)$-curve practically coincides with the curve corresponding to the bulk transition temperature, $T_{\rm IN}(x)$. However, at smaller pore sizes, particularly for $R \le$ 10 nm, the corresponding curves are shifted down. The smaller the pore radius, the more pronounced is the shift observed. This can also be seen in Fig.~4(b), where the $R^{-1}-$dependence of $T_{\rm PN}$ for a series of binary mixtures, i.e. at fixed mole fractions $x$, is explored.

While bulk properties of the nematic LCs, and particularly the molecular ordering at the isotropic-to-nematic phase transition, can be well described by a Landau-de Gennes theory, the KKLZ-model extends this phenomenological approach towards spatially confined nematic phases. This ansatz obviously depends on the specifics of the anchoring conditions at the channel walls. For the untreated alumina or silica surfaces, as employed in the present work, the walls of the channels enforce planar anchoring, whereas their elongated cylindrical geometry provides a preferred orientation of host rod-like molecules along the long channel axes. Following the KKLZ approach the orientational ordering is characterized by the scaled order parameter $q=Q/Q_0$, where $Q_0$  stands for the nematic order parameter value at the phase transition temperature of the unconfined system, i.e. $Q_0=Q(T_{\rm IN})$ \cite{Kutnjak,Kutnjak1}. The dimensionless free energy of the nematic LC is then represented as follows:
\begin{equation}
f=tq^2-2q^3+q^4-q\sigma+\kappa q^2 \label{eq1}
\end{equation}
where $t=(T-T^*)/(T_{IN}-T^*)$ is the dimensionless reduced temperature and $T^*$ is the effective temperature. A bilinear coupling between the order parameter and the nematic ordering field, i.e. the $q\sigma$-term, lowers the free energy of the nematic state. Particularly, it provides an upward shift of the effective transition temperature and causes a residual nematic ordering at high temperatures (paranematic state).  The $\kappa$-term, on the other hand,  attributable to quenched disorder, rises the free energy. Hence it results in a downward shift of the effective transition temperature.  Accordingly, the temperature of the nematic-to-paranematic transition is set by a competition of these two opposite contributions. Minimization of (\ref{eq1}) with respect to $q$ gives the equilibrium value of the scaled order parameter, $q_e$.  Its behavior is sketched in Fig.~5(a). Here curve 1 represents the molecular ordering in the bulk state ($\sigma=0,\kappa=0$). Curves 2-4 explore nematic-to-paranematic behavior at $\kappa=0$ in subcritical ($\sigma<\sigma_c$), critical ($\sigma=\sigma_c$) and overcritical ($\sigma>\sigma_c$) regimes, respectively, where $\sigma_c=0.5$. Curve 4$'$ represents overcritical behavior (curve 4) plus quenched disorder ($\kappa \ne 0$). To sum up, the nematic-to-paranematic transition occurs at $t_n=1+\sigma-\kappa$ and is discontinuous as long as $\sigma \le \sigma_c$. Above this value a continuous transition takes place.

\begin{figure}[tbp]
\begin{center}
  \epsfig{file=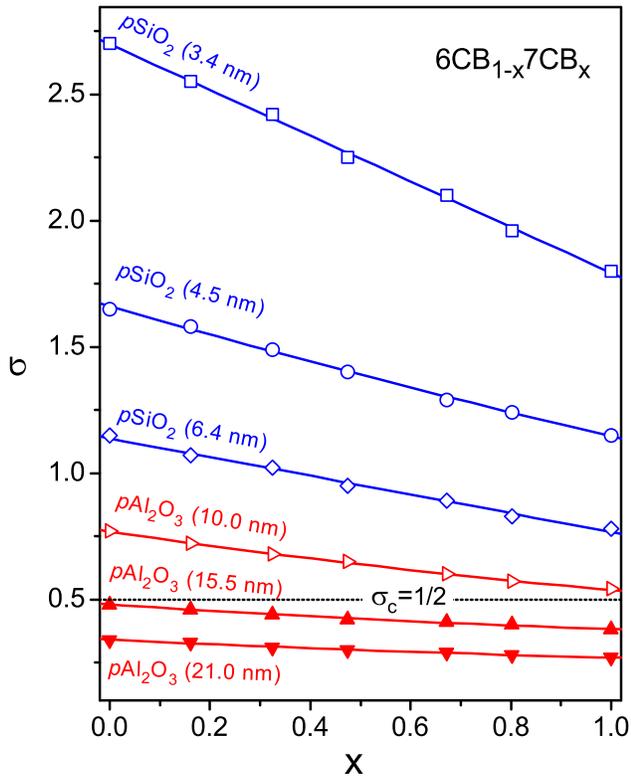, angle=0, width=0.99\columnwidth}
  \caption{\footnotesize (color online). Nematic ordering field $\sigma$ vs mole fraction $x$ as determined from the analysis (KKLZ-approach) of the measured birefringence of the binary nematic mixtures 6CB$_{1-x}$7CB$_x$ imbibed into alumina (red online color) and silica (blue online color) mesoporous membranes, see also labeled membrane type and corresponding channel radius in the brackets. The horizontal dot-line (black online color) marks the critical value of the nematic ordering field ($\sigma_c =1/2$) and separates the $\sigma-x$ plane in two regions: subcritical ($\sigma <\sigma_c$, solid symbols) and overcritical ($\sigma >\sigma_c$, open symbols).} \label{fig6}
\end{center}
\end{figure}

\begin{figure}[tbp]
\begin{center}
  \epsfig{file=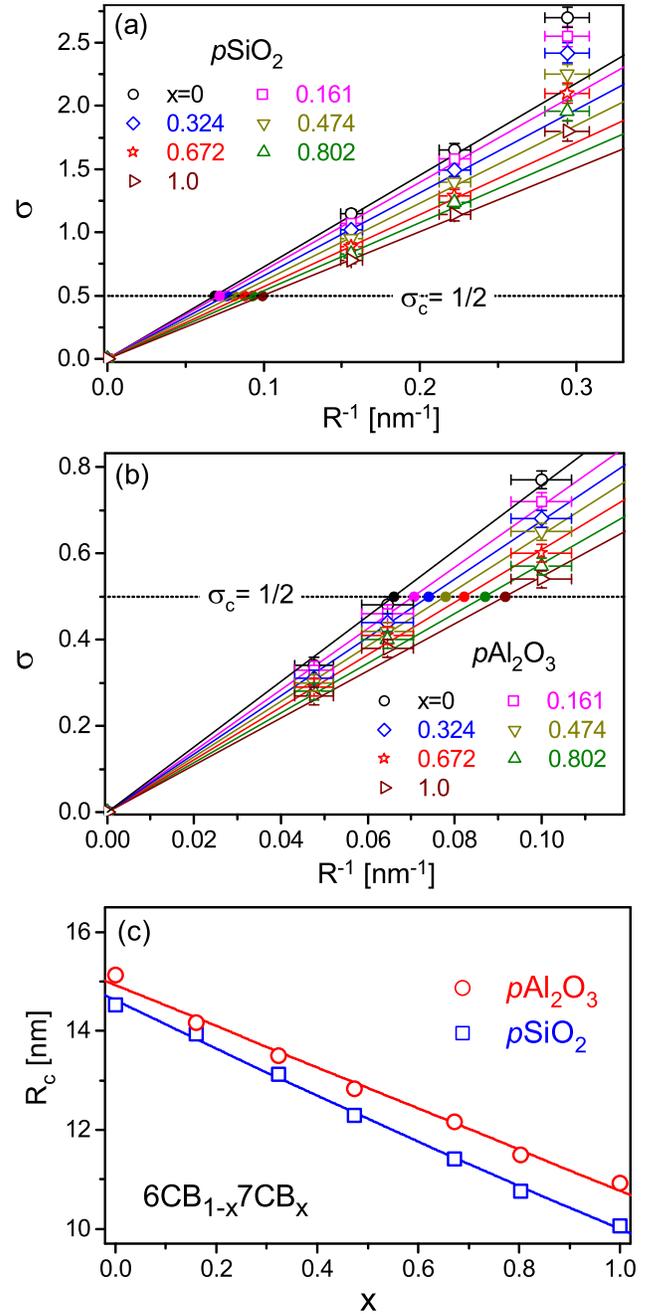, angle=0, width=0.99\columnwidth}
  \caption{\footnotesize (color online). The nematic ordering field $\sigma$ plotted as a function of $R^{-1}$ for a selected set of binary nematic mixtures 6CB$_{1-x}$7CB$_x$ (see $x$ labels) embedded into porous silica [panel (a)] and alumina [panel (b)] membranes. Symbols (plus error bars) are the data points extracted from the experiment. Lines are the best linear fits. Their intersection with the horizontal black dot line $\sigma=\sigma_c=1/2$ (see solid circles) yields the inverse critical radius, $R_c^{-1}$, as discussed in the text.  In panel (c) $R_c$ vs $x$ is depicted for binary nematic mixtures 6CB$_{1-x}$7CB$_x$ embedded into silica (squares, blue online color) and alumina (circles, red online color) membranes.} \label{fig7}
\end{center}
\end{figure}

Panels (b)-(i) of Fig.~5 present an analysis, where the KKLZ-approach is applied to the  measured birefringence of Fig.~3. For clarity, particularly in order to avoid an overlap of curves, the measured $\Delta n(T)$-dependences (open symbols of black, blue or green colors) likewise their best fits (solid lines, red online color) is depicted for selected systems: LCs 6CB ($x=$0), 7CB ($x=1.0$) and just one binary mixture, 6CB$_{1-x}$7CB$_x$ ($x=$0.474). The full set of extracted fitting parameters, i.e. for all the mole fractions $x$ studied in the present work, is listed in Table~1. The fitting procedure is based on an iterative, numerical minimization of Eq.(1) to extract $\sigma$- and $\kappa$-values, resulting in smallest least-square deviations between the calculated curve and the experimental points. The base lines corresponding to the thermooptic contribution (dash-dot line, gray online color) have been obtained within the same fitting procedure and are solely presented (as an example) for 6CB in each panel.

A crucial parameter of the model is the difference $T_{\rm IN}-T^*=\Delta T^*$, which calibrates the temperature scale of the KKLZ-model. Given the proximity of the system to the tricritical point, $\Delta T^*$ may, in principle, vary depending on the confinement conditions. Particularly it appears to be sensitive to surface wetting interactions \cite{Kralj10,Kralj11}. However, for relatively short and rigid nematogen molecules, such as e.g. 6CB or 7CB, the surface wetting is good \cite{Herminghaus}. Hence the corresponding confinement effects on $\Delta T^*$ are expected to be weak. It is ignored in the following.

In a first rough evaluation $\Delta T^*$ can be set as a free fit-parameter of the model as it has been done e.g. in Ref.\cite{Kityk2008}. In a more accurate analysis, however, it can be determined by analyzing the bulk birefringence in the region of the isotropic-to-nematic transition \cite{Kutnjak,Calus}. We follow here the second route by fitting the measured bulk birefringence. The fits based on a free energy expansion (1) [$\sigma=0,\kappa=0$], see solid red lines in Panel (b) of Fig.~5, provide a good description of the bulk birefringence mainly in the region of the nematic-to-isotropic transition, namely up to few degrees below $T_{\rm IN}$. Far below the transition the discrepancy between the experiment and theory is evident. The simple free energy expansion (equation (1)), which is limited by terms up to fourth-order in $q$ obviously cannot properly describe the order parameter saturation which takes place at lower temperatures. Including higher order terms (h.o.t.) into Eq.(1), up to fourteenth order, well reproduces the saturation of the nematic order parameter in the entire temperature region, see dash magenta lines in Fig.~5(b), but at the same time these higher terms do not much influence the $\Delta T^*$ fit-value. However, because of the large number of fit parameters in this case the extracted set of the free energy expansion coefficients appears to be ambiguous. Accordingly, we restrict our analysis to the free energy expansion of Eq.(1), trying to get the best fits in the region of the nematic-to-isotropic (paranematic) transition and extract fit values for $\Delta T^*,\sigma$, and $\kappa$.

The fitting of the measured bulk birefringence gives $\Delta T^*$ values which for the nematic binary mixtures 6CB$_{1-x}$7CB$_x$ vary practically linearly with $x$ between 3.2 K (6CB, $x=0$) and 4.9 K (7CB, $x=1.0$), see Table~1. It means that the first order character of the nematic-to-isotropic transition somewhat strengthens while $x$ changes from 0 to 1. The order parameter saturation is also observable for the LCs 6CB, 7CB and for their binary mixtures confined into nanochannels. Although it is evident specifically for the pores of large radii [compare e.g. panels (c),(e) or (g) and (d),(f) or(i)]. This means that the order parameter saturation originates dominantly from the molecular ordering in the vicinity of the pore axis, i.e. in the core region of the channel filling. By contrast, the nematic ordering near the channel walls appears far from saturation,  presumably caused by the host-guest interactions and by wall roughness enforcing molecular disorder. In other words, for large  pore radii the effective nematic order is favored by the ordered core component,  whereas at small ones the disordered near-interface layers provide a dominant contribution.

Insignificant deviations between the experimental points and the fitting curves can also be found directly in the transition region. In a number of cases the fits exhibit slightly steeper changes than observed in the experiment. These discrepancies presumably originate in wall roughness and/or pore diameters distributions, which are ignored in the simple analysis presented here.  A consideration of them would lead to a more smeared transition behavior.

\begin{figure}[tbp]
\begin{center}
  \epsfig{file=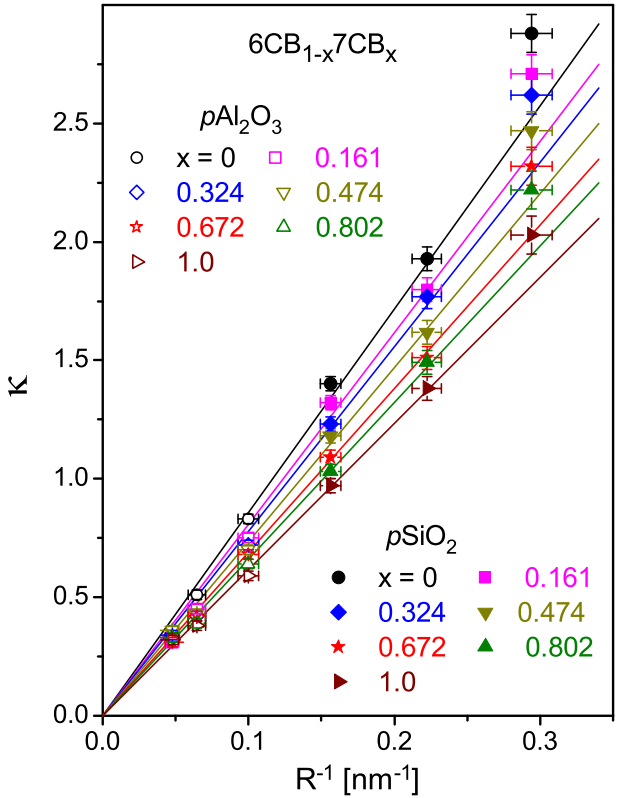, angle=0, width=0.99\columnwidth}
  \caption{\footnotesize (color online).  Strength of quenched disorder, $\kappa$ plotted as a function of $R^{-1}$ for a series of the binary nematic mixtures 6CB$_{1-x}$7CB$_x$ (see $x$ labels) embedded into porous silica (solid symbols supplied by error bars) and alumina (open symbols supplied by error bars) membranes. Lines are the best linear fits.} \label{fig8}
\end{center}
\end{figure}

\begin{table*}[ht]
\footnotesize
 \vspace{0.5cm}
\caption{The set of fitting parameters, $\Delta T^*,\sigma$ and $\kappa$ as extracted from the phenomenological analysis (KKLZ-approach) employed to the measured optical birefringence for binary nematic mixtures 6CB$_{1-x}$7CB$_x$ in the bulk state and confined into porous alumina ($p$Al$_2$O$_3$) and silica ($p$SiO$_2$) membranes.}
\vspace{0.5cm} 
\begin{tabular}{|c|c|c|c|c|c|c|c|c|c|c|c|c|c|} 
\hline\hline 
\cline{2-14}
&bulk cell&\multicolumn{6}{c}{$p$Al$_2$O$_3$}\vline&\multicolumn{6}{c}{$p$SiO$_2$}\vline\\
\cline{2-14}
\small 6CB$_{1-x}$7CB$_x$& $h=$20 $\mu$m &\multicolumn{2}{c}{$R=21.0$ nm}\vline &\multicolumn{2}{c}{$R=15.5$ nm}\vline &\multicolumn{2}{c} {$R=10.0$ nm}\vline & \multicolumn{2}{c}{$R=6.4$ nm}\vline & \multicolumn{2}{c}{$R=4.5$ nm} \vline&\multicolumn{2}{c}{$R=3.4$ nm}\vline \\ [0.5ex] 
\cline{2-14} 
$x$&$\Delta T^*$ [K]&$\sigma$&$\kappa$&$\sigma$&$\kappa$&$\sigma$&$\kappa$&$\sigma$&$\kappa$&$\sigma$&$\kappa$&$\sigma$&$\kappa$\\
\hline
0.0&3.2&0.34&0.33&0.48&0.51&0.77&0.83&1.15&1.40&1.65&1.93&2.70&2.88\\
0.161&3.5&0.33&0.31&0.46&0.45&0.72&0.75&1.08&1.32&1.58&1.80&2.55&2.71\\
0.324&3.8&0.31&0.34&0.44&0.42&0.68&0.72&1.02&1.23&1.49&1.77&2.42&2.62\\
0.474&4.1&0.30&0.36&0.42&0.43&0.65&0.7&0.95&1.18&1.40&1.62&2.25&2.47\\
0.672&4.4&0.29&0.31&0.41&0.42&0.60&0.68&0.89&1.09&1.29&1.51&2.10&2.32\\
0.802&4.6&0.28&0.33&0.40&0.39&0.57&0.64&0.83&1.03&1.24&1.49&1.96&2.22\\
1.0&4.9&0.27&0.32&0.38&0.38&0.54&0.59&0.78&0.97&1.14&1.38&1.80&2.03\\
\hline 
\end{tabular}
\label{tb1}
\end{table*}

In Fig.~6 the nematic ordering field $\sigma$ is plotted versus the mole fraction $x$ as determined by analyzing the measured birefringence within the KKLZ-approach for the nematic binary mixtures 6CB$_{1-x}$7CB$_x$ ($0\le x\le1$) embedded into nanoporous membranes of alumina (red online color) and silica (blue online color). Here, the horizontal doted line (black online color) marks the critical value of the nematic ordering field ($\sigma_c =1/2$). It separates the $\sigma-x$ plane in two regions: subcritical ($\sigma <\sigma_c$, solid symbols) and overcritical ($\sigma >\sigma_c$, open symbols). Whereas the overcritical behavior for the nematic order parameter has been proven in several recent studies on nCB embedded into silica and silicon membranes\cite{Kityk2008,Kityk1,Calus}, the present work documents examples of subcritical behavior for the nematogen LCs 6CB and 7CB as well as their binary mixtures embedded into the alumina membranes of large pore radii ($R>15$ nm). For both types of the membranes and all the pore radii investigated $\sigma$ exhibits a nearly linear decrease for a mole fraction $x$ change from 0 to 1.

Following the Refs.\cite{Kutnjak,Kutnjak1} the nematic ordering field in the case of parallel axial ordering is expressed as:
\begin{equation}
\sigma=\frac{2\xi^{*2}W_{n1}}{RkQ_0} \label{eq2}
\end{equation}
where $\xi^*=\xi(T_{\rm IN})$ is the correlation length at the temperature of the isotropic-to-nematic transition, $W_{n1}$ is the anchoring strength and $k$ is the nematic elastic constant. The $x$-dependence of $\sigma$ originates from varying material constants characterizing the nematic binary mixtures as well as anchoring forces. According to Eq.(2) $\sigma \propto R^{-1}$. To verify this scaling $\sigma$ vs $R^{-1}$ is shown in Fig.~7 for a series of the nematic binary mixtures 6CB$_{1-x}$7CB$_x$ embedded into porous silica [panel (a)] and alumina [panel (b)] membranes. In the case of the alumina membranes for each mole fraction $x$ all three experimental data points can be reasonably well fitted (within the error bars) by a linear dependence.

This is not the case, however, for the silica membranes. Particularly, the data points corresponding to the smallest pore radius ($R=3.4$ nm),  evidently deviate  from the linear fits drawn through the two other data points, $R=6.4$ nm and $R=4.5$ nm. This deviation, which well exceeds the experimental error, could originate in several reasons, most probably in surface roughness of the pore walls, a nonuniform pore diameter along the channels and a nonuniform size of different pores (pore size distribution). The first reason leads to a decrease in the effective nematic ordering, mainly in the interface region, as has been outlined in Ref.\cite{Calus}. Accordingly, Eq.(2), which is derived considering smooth pore walls, holds as long as the number of molecules located in the interfacial region remains negligible compared to the number in the core. The second and third reasons, which are related directly to a pore size variation, result to a shift of the "local" transition temperature in accordance with a local channel diameter. The distribution of pore sizes is thus transferred to a distribution of local transition temperatures, which "smears" the behavior of the effective (average) order parameter in the phase transition region \emph{additionally} to what is expected within the KKLZ approach. This effect becomes especially crucial for small pore radii, if one assumes that the  absolute deviations from the mean pore diameter is of similar magnitude for all membranes.  The applicability of Eq.(2) is therefore limited to silica membranes with pore radii larger than roughly 4 nm. Obviously, this value may vary depending on the specifics of the host-guest interaction, the pore wall roughness and the pore size distribution.

Within the KKLZ ansatz the slope of the $\sigma(R^{-1})$-dependence is given by the factor $\varrho=2\xi^{*2}W_{n1}/(kQ_0)$. The fit of the data points in Fig.~7(a) and 7(b) yields $\varrho$-factor which varies from 5 to 7.5 nm. With the typical correlation length $\xi^*\sim$10-15 nm, elastic constant $k$ =$(8.0\div9.0)\cdot$10$^{-12}$ N, \cite{elastic} and $Q_0\sim$1/2 one yields an anchoring strength $W_{n1}$ in the range 50-100 $\mu$J/m$^2$ which corresponds to a regime of weak anchoring.

\begin{figure}[tbp]
\begin{center}
  \epsfig{file=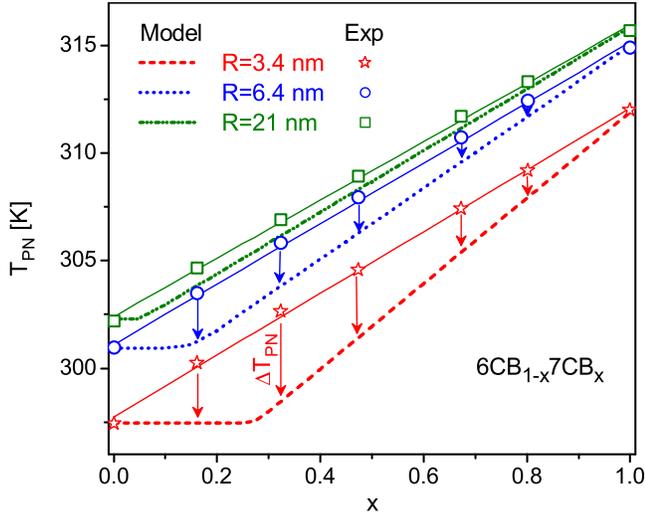, angle=0, width=0.99\columnwidth}
  \caption{\footnotesize (color online). $T_{\rm PN}$ vs $x$ for binary nematic mixtures 6CB$_{1-x}$7CB$_x$ confined into nanochannels for a selected set of channel diameters as indicated in the figure. The symbols are the experimental data points. The solid lines trace $T_{\rm PN}(x)$-dependences as expected for spatially homogeneous mixtures, i.e. with no separation of the components. Broken curves explore hypothetical $T_{\rm PN}(x)$-dependences assuming a separation of the components caused by a strong preferential adsorption of 7CB molecules on the pore walls by interfacial interactions. Downward arrows sketch the shift of the transition point, $\Delta T_{\rm PN}$ resulting from an hypothetical component separation in the confined state. The thickness of the first molecular layer is assumed to be 5 \AA \quad in this calculation.} \label{fig9}
\end{center}
\end{figure}

The intersection of the linear fits with the horizontal black dot line $\sigma=\sigma_c=1/2$ (see solid circles) yields the inverse critical radii, $R_c^{-1}$. In Fig.~7(c)$R_c$ vs $x$ for nematic binary mixtures 6CB$_{1-x}$7CB$_x$ confined into silica (squares, blue online color) and alumina (circles, red online color) membranes is shown. For 6CB ($x=0$) embedded into silica membranes  $R_c=$14.3 nm, i.e within the experimental error ($\pm$0.5 nm) it well agrees with the $R_c$-value of 14.0 nm obtained in recent studies \cite{Calus} on silicon and silica membranes. For 7CB ($x=$1.0), on the other hand, the value of the critical radius value is 10.0 nm, somewhat larger than its magnitude $R_c=$8.0 nm reported in our earlier studies \cite{Kityk2008}. Note, however, that the $R_c$-value in Ref.\cite{Kityk2008} relies on an analysis neglecting possible $\Delta T^*$ and hence is less reliable than the value obtained here.

The critical radii, $R_c$ of the silica and alumina membranes differ marginally, only, if one considers the experimental error. This implies nearly identical surface wetting strengths for the silica and alumina membranes. This may appear surprising at first glance. The oxide surfaces (SiO$_2$ or Al$_2$O$_3$) are covered, however, in both cases by a hydrophilic monomolecular layer consisting of hydroxyl (OH) groups \cite{Zitat1,Zitat2} obviously resulting in similar wetting and anchoring conditions.

The quenched disorder effect, quantified by the $\kappa$-value, exhibits also a simple scaling behavior. In Fig.~8 $\kappa$ vs $R^{-1}$ is plotted for a series of the nematic binary mixtures 6CB$_{1-x}$7CB$_x$ embedded into porous silica (solid symbols supplied by error bars) and alumina (open symbols supplied by error bars) membranes. Provided one excludes the data points corresponding to the membrane with smallest pore radius ($R=$3.4 nm) the remaining measurements can well be fitted by a linear behavior.

The $\kappa$-value, similarly as the nematic ordering field, $\sigma$, exhibits the same type of scaling behavior, i.e. $\propto R^{-1}$, in good agreement with a simple consideration: If one takes into account that quenched disorder effects have interfacial origin, the corresponding extra contribution to the free energy is proportional to the wall surface. This energy contribution scales $\propto R$ for a cylindrical geometry. The remaining free energy contribution, originating from the core region, scales $\propto R^2$. Hence, the relative weight of the quenched disorder effects, quantified by the magnitude of $\kappa$, should scale like the ratio of surface to volume, i.e. $\kappa \propto R^{-1}$. This scaling is obeyed for sufficiently large pore diameters only, i.e. when the weight of the surface contribution remains still considerably smaller than the volume one. Our experiment shows that it is valid here for pore radii roughly larger than $\sim$4~nm.

Finally, we return to a  principal issue regarding possible molecular arrangements of the nematic mixture in the cylindrical nanochannels. We try now to answer the question outlined in Fig.~1, i.e. on whether a local stoichiometry of mixture remains constant over the pore volume or is subjected to a separation of the components due to possible differences in anchoring (and or attractive interfacial) forces.

Let's assume that the molecular layer next to the pore wall after thermodynamic equilibration is indeed occupied preferably by the molecules of a certain type. This automatically leads to a reduction of their number in the core region modifying thus the local stoichiometry and hence its properties. It is well established that the monomolecular surface layer next to the pore wall does not support the nematic-to-paranematic transition \cite{Huber2}. It means that the phase transition point is basically set by the properties of the core region. Since its content has been modified due to the separation of the components, the effective transition temperature should shift accordingly.

How large would this effect be? In our evaluation we rely on intermolecular spacings $a_0$ of about 5 \AA, reported in Ref.\cite{Komolkin} for 5CB, which is expected to be valid also for 6CB and 7CB. We assume that the thickness of the monomolecular layer next to the pore wall roughly corresponds to the intermolecular distance $a_0$ resulting in a volume fraction of this layer in a total pore volume, $x_m=1-(1-a_0/R)^2$. Considering a binary mixture of 6CB$_{1-x}$7CB$_{x}$ and assuming separation of the components due to a strong preferential adsorption on the pore walls molecules of a certain type (say e.g. 7CB) their fraction in the core region, $x_c$, equals 0 as long as $x \le x_m$, otherwise it is defined as $x_c=(x-x_m)/(1-x_m)$. The $x$-dependence of the bulk isotropic-to-nematic transition can be well approximated by a linear dependence as $T_{\rm IN}(x)=T_{\rm IN}(x=0)+x[T_{\rm IN}(x=1.0)-T_{\rm IN}(x=0)]$. Translating this to the confined state, the effective transition temperature $T_{\rm PN}$ is defined by the fraction of 7CB molecules in the core region, $x_c(x)$, i.e. $T_{\rm PN}(x_c(x))=T_{\rm IN}(x=0)-\Delta_R+x_c(x)\cdot[T_{\rm IN}(x=1.0)-T_{\rm IN}(x=0)]$, where $\Delta_R$ is a $R$-dependent constant temperature onset (downward shift)  due to confinement effects.

The result of this consideration is sketched in Fig.~9. Here symbols are the experimental data points found for several pore radii, namely $R=$ 3.4 nm, 6.4 nm and 21.0 nm. Solid lines indicate $T_{\rm PN}(x)$-dependences typical of a spatially homogeneous mixture, i.e. with no separation of the components. Broken curves explore hypothetical $T_{\rm PN}(x)$-dependences assuming strong preferential adsorption of 7CB molecules on the pore walls. It is obvious that the effect due to separation of the components is expected to be strongly $R$-dependent. Particularly, it is negligible at large pore radii and vice versa. At small pore sizes,  $T_{\rm PN}(x)$  exhibits an evident nonlinear variation with a characteristic plateau corresponding to a constant $T_{\rm PN}$-value in the range, $0 \le x \le x_m$, and then it is fast rising for $x > x_m$.

If the component separation caused by interfacial interactions would exist in the nanoconfined mixtures, it would be easily detectable for the membranes with pore radii approx. less than 10 nm. This is not the case. Therefore, we can conclude that the binary mixtures of cyanobiphenyls 6CB and 7CB keeps its homogeneity of the original stoichiometry also in the nano confined state, at least for pores with diameters up to 7 nm. 

An indirect evidence follows also from the behavior of other properties, for instance from the nematic ordering field, $\sigma$. It is strongly dependent on the anchoring strength and elastic constants of the relevant LC species. Hence, it's linear variation with the molar fraction $x$ also corroborates the homogeneous state of the mixture.

We attribute the conserved homogeneity of the two components in the nano channels to the chemical similarity of the two components (and thus to the very similar intermolecular and molecule/pore wall interactions). Moreover, a demixing would lead to huge interfacial areas in the confined geometry, given the large specific surface areas of the mesoporous hosts. Therefore, the corresponding energetic costs of fluid/fluid interfaces and of isotropic-nematic interfaces (because of the distinct $T_IN$ of the single components) may be non negligible and additionally contribute to a suppression of any demixing, despite the quite small magnitude of the specific interfacial tensions of the liquid/liquid (a few mN/m) and of the isotropic-nematic interfaces (on the order of 10$^{-2}$mN/m) \cite{Faetti1984, Faetti1985}.

Given the distinct isotropic-to-nematic phase transition temperatures of the single components, in the case of a chemical segregation a coexistence of a nematic phase (of the one component) with an isotropic phase (of the second component) and isotropic/nematic interfaces would occur. The corresponding interfacial tensions are on the order of 10$^{-2}$mN/m for cyanobiphenyl-based mesogens, only. These excess energies are avoided for the homogeneous systems observed here, which may contribute to the good miscibility observed.    
\section{4. Conclusion}
To the best of our knowledge we have reported here the first experimental study of the nematic ordering of \textit{binary} mixtures embedded in nanochannels. The results are compared with the bulk behavior and subjected to a phenomenological analysis. The molecular orientational order inside the channels gives rise to an excess birefringence, which is proportional to the nematic order parameter and appears on the background of the geometric birefringence typical of porous matrices with parallel-aligned nanochannels. By applying a fitting analysis based on a Landau-De-Gennes model for cylindrical confinement (Kutnjak-Kralj-Lahajnar-Zumer approach), we were able to separate these distinct birefringence contributions and thus to determine the magnitude of the effective nematic ordering field, $\sigma$ originating from the geometric confinement. Moreover, we were able to characterise quantitatively quenched disorder effects (the $\kappa$-term of the dimensionless free energy of the KKLZ model).

Depending on the pore radius the nematic order in the cylindrical nanochannels evolves either discontinuously (subcritical regime, $\sigma<1/2$) or continuously (overcritical regime, $\sigma>1/2$), but in both cases  with a characteristic precursor behavior typical of a paranematic state. Whereas the overcritical behavior for the nematic order parameter has been documented in several previous optical polarimetry studies on nCBs embedded into silica and/or silicon membranes\cite{Kityk2008,Kityk1,Calus}, we present here evidence for its subcritical behavior for the binary mixtures embedded into the alumina membranes with pore radii larger than 15 nm.

The strength of the ordering fields, imposed by the pore walls, appears in all cases as dependent on the pore radius, $R$, and molar fraction, $x$. It varies linearly with $x$, and scales as $R^{-1}$ for the pore radii larger than approx. 4 nm. For smaller pore radii, e.g. $R=$3.4 nm, we found hints of deviations of the $R^{-1}$-$\sigma$ scaling, which we attribute to the surface roughness of the pore walls, non-uniform pore diameters along the channels' long axes and/or a significant pore size distribution. An interpolation of the $\sigma (R^{-1})$-dependence in the linear regime provides values for the critical pore radius, $R_c$, separating continuous from discontinuous behavior. It varies linearly with $x$ and differs negligibly for the silica and alumina membranes. We suggest that the native oxide surfaces are covered in both cases by hydrophilic monomolecular layers consisting of hydroxyl (OH) groups, which results in  practically identical tangential anchoring strengths.

Quenched disorder effects, described by the $\kappa$-term of the dimensionless KKLZ free energy, exhibits the same type of scaling behavior as $\sigma$, i.e. $\propto R^{-1}$. This scaling is rationalized by the interfacial origin of this disorder contribution. Again, this scaling is valid for pore radii larger than approx. 4 nm.

Linear variations of the nematic-to-paranematic transition point $T_{\rm PN}$ and the nematic ordering field, $\sigma$ vs the mole fraction $x$ suggest that the binary mixtures of cyanobiphenyls 6CB and 7CB keep their homogeneous stoichiometry also upon nanoconfinement, at least for channels with diameters larger than 7 nm. This allowed us to successfully apply the simple phenomenological models outlined above. 

In general, we believe our findings are not only of fundamental interest, but also of significant importance for nanotechnological applications. They allow for predictable and easily tunable properties of liquid crystalline systems at the nanoscale, similarly as it has been extensively employed for decades in applications of binary LCs.

For the future we envision an exploration of alternative pure LC constituents. For less chemically related components than investigated here, the observation of inhomogeneous mixtures in the presence of a preferred adsorption of one LC is very well imaginable and may lead to additional, interesting new phenomenologies, similarly as it has been found for simple liquids confined in pores \cite{Gelb1999,Woywod2003, Woywod2005, Rother2004, Schemmel2005,Lefort2011}. Moreover, the interpretation of the present experiments, in particular for the small channel diameters (below 4 nm) could profit from detailed molecular dynamics simulations \cite{MD}.

\section{Acknowledgement}

This work has been supported by the Polish National Science Centre under the Project "Molecular Structure and Dynamics of Liquid Crystals Based Nanocomposites" (Decision No. DEC-2012/05/B/ST3/02782). P.H. acknowledges support from the German Research Foundation (DFG) through SFB 986 ÓTailor- Made Multi-Scale Materials Systems M3Ó, Hamburg (Germany).

\end{document}